\documentclass[prb,aps,showpacs,floatfix,twocolumn,superscriptaddress]{revtex4-1}

\usepackage{graphicx}
\usepackage{amssymb}
\usepackage{color}
\usepackage{amsmath} 
\newcommand{\be}{\begin{eqnarray}}
\newcommand{\ee}{\end{eqnarray}}

\begin{document}
\title{Phase transitions in the frustrated Ising model on the square lattice}

\author{Songbo Jin}
\affiliation{Department of Physics, Boston University, 590 Commonwealth Avenue, Boston, Massachusetts 02215, USA}

\author{Arnab Sen}
\affiliation{Max-Planck-Institut f\"{u}r Physik Komplexer Systeme, 01187 Dresden, Germany}

\author{Wenan Guo}
\affiliation{Department of Physics, Beijing Normal University, Beijing, 100875, China}

\author{Anders~W.~Sandvik}
\affiliation{Department of Physics, Boston University, 590 Commonwealth Avenue, Boston, Massachusetts 02215, USA}
\date{\today}

\begin{abstract}
We consider the thermal phase transition from a paramagnetic to stripe-antiferromagnetic phase in the frustrated two-dimensional square-lattice 
Ising model with competing interactions $J_1<0$ (nearest neighbor, ferromagnetic) and $J_2 >0$ (second neighbor, antiferromagnetic). The striped 
phase breaks a $Z_4$ symmetry and is stabilized at low temperatures for $g=J_2/|J_1|>1/2$. Despite the simplicity of the model, it has proved difficult to 
precisely determine the order and the universality class of the phase transitions. This was done convincingly only recently by Jin {\it et al.} [PRL {\bf 108}, 
045702 (2012)]. Here, we further elucidate the nature of these transitions and their anomalies by employing a combination of cluster mean-field theory, 
Monte Carlo simulations, and transfer-matrix calculations. The $J_1$-$J_2$ model has a line of very weak first-order phase transitions in the whole 
region $1/2<g<g^*$, where $g^* = 0.67 \pm 0.01$. Thereafter, the transitions from $g=g^*$ to $g \rightarrow \infty$ are continuous and can be fully mapped, 
using universality arguments, to the critical line of the well known Ashkin-Teller model from its $4$-state Potts point  to the decoupled Ising limit. 
We also comment on the {\em pseudo-first-order} behavior at the Potts point and its neighborhood in the Ashkin-Teller model on finite lattices, 
which in turn leads to the appearance of similar effects in the vicinity of the multicritical point $g^*$ in the $J_1$-$J_2$ model. The continuous 
transitions near $g^*$ can therefore be mistaken to be first-order transitions, and this realization was the key to understanding the paramagnetic-striped
transition for the full range of $g>1/2$. Most of our results are based on Monte Carlo calculations, while the cluster mean-field and transfer-matrix
results provide useful methodological bench-marks for weakly first-order behaviors and Ashkin-Teller criticality.

\end{abstract}


\maketitle

\section{Introduction}

\begin{figure}
\center{\includegraphics[width=5cm, clip]{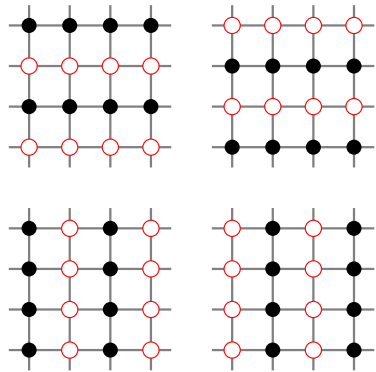}}
\vskip-1mm
\caption{(Color online) The four symmetry related striped ground states of the $J_1$-$J_2$ model when $g>1/2$. The striped phase breaks a $Z_4$ symmetry.
Solid and open circles represent the spin states $\sigma_i = \pm 1$.}
\label{stripes}
\vskip-3mm
\end{figure}

The Ising model with nearest-neighbor interactions on the two-dimensional (2D) square lattice presents a rare instance where the partition function can be computed 
exactly at any temperature $T$.~\cite{Onsager44} This allows for the calculation of the critical exponents characterizing the continuous phase transition between the 
magnetically ordered ferromagnet and the disordered paramagnetic state. Adding competing (frustrated) interactions provides a route for the appearance of new phases 
and, in some cases, new types of phase transitions outside the Ising universality class. A next-nearest-neighbor antiferromagnetic interaction represents the 
simplest way to incorporate frustration in the standard Ising model. This model, the $J_1$-$J_2$ Ising model, is defined by the Hamiltonian
\be
H = J_1\sum_{\langle ij \rangle}\sigma_i \sigma_j + J_2 \sum_{\langle \langle ij \rangle \rangle}\sigma_i \sigma_j,
\label{eq1}
\ee
where first and second (diagonal) neighbors on the square lattice are denoted by $\langle ij \rangle$ and $\langle \langle ij \rangle \rangle$, respectively, and
$\sigma_i = \pm 1$. When the ratio $g = J_2/|J_1|<1/2$, there is an Ising transition versus $T$ to a ferromagnetic state.~\cite{Nightingale,Swendsen,Oitmaa,Binder,Landau} 
The competing interactions in the model stabilize a new striped phase (see Fig.~\ref{stripes}) when $g > 1/2$. Since these stripes can be oriented in either the $x$ or 
the $y$ lattice direction, the ordering breaks a four-fold ($Z_4$) symmetry on the square lattice. Increasing the temperature from $T=0$ at a fixed $g >1/2$, a transition 
to a disordered state occurs with no other intermediate broken symmetry phase in between. In this paper we study the phase transition into the striped state.

Unlike the Ising transition to a ($Z_2$ ordered) ferromagnetic state, the nature of the phase transition between a $Z_4$ ordered state and a disordered
state in 2D cannot be determined simply from the symmetry of the order parameter. This is an example of weak universality, a concept first introduced by 
Suzuki,\cite{Suzuki} where the dimensionality of the system and the symmetry properties of the order parameter are not enough to fix the universality and hence, 
the critical exponents of the phase transition. The exponents may vary with some tuning parameter in the system even though the symmetry of the order parameter 
does not change. Only certain ratios of the critical exponents remain fixed and these define \cite{Suzuki} the weaker form of universality (or, equivalently,
the exponent $\eta$ describing the correlation function at the critical point is fixed, while other exponents vary). Some exotic 2D models, where the critical 
exponents can be analytically calculated as a function of a coupling parameter, indeed show this behavior, e.g., the eight-vertex model~\cite{eightv} and the
Ashkin-Teller (AT) model.~\cite{ATor, Nienhuis, AT}

Even though the frustrated $J_1$-$J_2$ model defined by Eq.~(\ref{eq1}) represents perhaps the simplest generalized 2D Ising model, its stripe transition 
remained highly controversial until recently, despite several past 
studies.\cite{Nightingale,Swendsen,Oitmaa,Binder,Landau,LandauBinder,variational1,variational2,honecker1,honecker2} 
Early numerical and analytic approaches supported the idea that the transition is always continuous for $g>1/2$, but with critical exponents that vary with 
$g$, thus providing an example of weak universality. However, some variational studies~\cite{variational1,variational2} and recent Monte Carlo (MC) 
studies \cite{honecker1,honecker2} have found a line of first-order transitions for $1/2<g\lesssim 1$. 

One recent MC study by Kaltz et al.~\cite{honecker2} 
used the existence of a double-peak structure in energy histograms to conclude that the transition is first-order up to $g=g^*$, with $g* \approx 0.9$. For 
higher $g$, in the same work a continuum field theory was derived perturbatively in $1/g$, resulting in an AT-like model. For intermediate values of $g$, where 
the (perturbative) field theory cannot be expected to be reliable, and the MC results were ambiguous, it was not possible to definitely conclude that the AT 
scenario holds all the way down to $g^*$. In particular, deviations from $\eta=1/4$ (the fixed value of this exponent in the AT model) were seen for $g$ in the range $1-5$. 

In another recent study,~\cite{prl12} it was shown by three of the present authors that the stripe transition  is first order in a much smaller range of couplings
than previously believed; for $1/2<g<g^*$, with $g^* \approx 0.67$. For $g>g^*$ it is continuous and in the AT universality class. The exponents change continuously with 
$g$ as in the AT model~\cite{AT}, with $g^*$ corresponding to the universality of the $4$-state Potts model~\cite{Baxter,Salas} (which is equivalent to the AT model at 
one end-point of a critical line) and $g \rightarrow \infty$ to standard Ising universality. While AT criticality had been suspected at the stripe transition earlier, 
it had not been possible to demonstrate this convincingly for a wide range of couplings before. The key to solving this problem was the realization that the Potts 
model harbors {\it pseudo-first-order} behavior and (previously known\cite{Salas}) logarithmic corrections, and that these match very well the properties of the 
$J_1$-$J_2$ model at $g\approx 0.67$. Thus, the full critical curve bridging the Ising and $4$-state Potts point of the symmetric version of the AT model (which 
we will define in detail further below) can be completely realized in the $J_1$-$J_2$ model.

The pseudo-first-order behavior found in Ref.~\onlinecite{prl12} implies that indicators (necessary but not sufficient conditions) of first-order transitions, e.g., multiple 
peaks in energy and order-parameter distributions, lead to over-estimation of the region of discontinuous transitions in this model. Mere observation of multi-peak structures
is not sufficient for concluding that a transition is first-order, but careful finite-size scaling studies are required to extrapolate, e.g., the latent heat to infinite size.
The $4$-state Potts model and neighboring transitions in the AT model exhibit clear pseudo-first-order behavior~\cite{prl12} for finite sizes, though these transitions are known 
to be continuous.~\cite{AT} It is then necessary to look at certain universal properties in the $J_1$-$J_2$ model to determine whether the transition is continuous and belongs 
to the AT universality class. This pseudo-critical behavior in the $J_1$-$J_2$ model was also verified recently in Ref.~\onlinecite{honecker3}, where the double-peak structure 
in the energy histogram was shown to disappear at large system sizes ($L \sim 2000$ for a periodic $L \times L$ system) for $g=0.80$ (while in Ref.~\onlinecite{prl12} the
order-parameter histograms were analyzed).  

In this article we present further evidence to support this picture~\cite{prl12, honecker3} of the transitions from the striped phase in the $J_1$-$J_2$ model.
In addition to MC simulations, we also consider cluster mean-field (CMF) and numerical transefer-matrix (TM) calculations. While in the end MC calculations appear 
to be the only way to reliably study the stripe transition close to the most interesting point $g=g^*$, due to the subtleties discussed above, it is still useful to
bench-mark these other commonly used methods.

The rest of the paper is organized in the following way: In Sec.~\ref{sec:universality} we briefly summarize the known scenarios for continuous phase transitions 
from a $Z_4$ ordered to a disordered phase in 2D. We then investigate the phase transitions of the $J_1$-$J_2$ model in detail using the cluster mean-field theory 
approach (Sec.~\ref{sec:CMF}), MC simulations (Sec.~\ref{sec:MC}) and the TM approach (Sec.~\ref{sec:TMF}). We also present some further results for the 
AT model in Sec.~\ref{sec:MC}, including its pseudo-first-order behaviour near the $4$-state Potts point. We further establish the equivalence between the continuous 
phase transitions in the AT and $J_1$-$J_2$ models, including quantitative results for how the parameters of the two models correspond to each other in terms 
of the varying critical indices. We give a brief summary of the results in Sec.~\ref{sec:summary}.

\section{Expectations from universality}
 \label{sec:universality}

In two dimensions, the critical exponents can have various possible values when the ordered phase breaks a $Z_4$ symmetry. In the $J_1$-$J_2$ model for $g \ge 1/2$, only 
the $g \rightarrow \infty$ limit and $g=1/2$ transitions are exactly known. At $g \rightarrow \infty$, the system consists of two decoupled Ising systems and there is a 
continuous thermal phase transition in the Ising universality class. At $g=1/2$, it is clear that there is a first-order transition at $T=0$ between a ferromagnetic 
and a stripe-antiferromagnetic state of the type depicted in Fig.~\ref{stripes}. The first-order transition point is unusual in that there is a co-existence of a large number 
of states~\cite{honecker1} made up entirely of horizontal (or, vertical) stripes where the orientation ($\sigma=+1$ or $-1$) of each stripe can be chosen independently. 
The nature of the $g>1/2$, $T>0$ transitions is not {\em a priori} clear. We briefly discuss two microscopic scenarios which cover the known theoretical possibilities 
for continuous transitions out of a $Z_4$ symmetry-broken 2D state.

Let us first consider the 2D XY model in a four-fold anisotropic field of strength $h_4$,
\be
H = -\sum_{\langle ij \rangle}\cos(\theta_i-\theta_j) -h_4 \sum_i \cos(4 \theta_i),
\ee
where the sites $i$ reside on a square lattice and $\theta_i$ defines a 2D fixed-length vector in the XY plane. At $h_4=0$, there is a Kosterlitz-Thouless (KT) 
transition versus temperature \cite{KT1,KT2} while $|h_4| \rightarrow \infty$ gives the standard Ising universality. A non-zero $h_4$ leads to a four-fold broken symmetry 
phase at low $T$. The critical exponents change as a function of $h_4$, e.g., the thermal exponent $\nu$ equals $1$ in the Ising limit and $\nu \rightarrow \infty$ at the  
KT transition. However, the important observation for our purpose here is that the specific heat exponent $\alpha/\nu$ pertaining to finite-size scaling equals $0$ in the 
Ising limit (the specific heat diverges logarithmically with system size here) and develops a cusp at finite $h_4$ indicating a negative $\alpha$.  This cannot possibly 
explain the behavior of the specific heat in the $J_1$-$J_2$ model~\cite{prl12} where the divergence with system size seems quite strong for all $g>1/2$, indicating 
$\alpha/\nu >0$ if the transition is assumed to be continuous.

Next, we consider the AT model on the square lattice~\cite{ATor, AT} which can be written as
\be
H = -\sum_{\langle ij \rangle}(\sigma_i \sigma_j + \tau_i \tau_j + K \sigma_i \sigma_j \tau_i \tau_j),
\label{ATmodel}
\ee 
where two Ising variables $\sigma_i, \tau_i$ reside on each site $i$ of the square lattice and are coupled to each other through $K$. There is a symmetry 
of the model, corresponding to the permutations of the variables $\sigma$, $\tau$, and $\sigma\tau$. These map the Hamiltonian (\ref{ATmodel}) onto itself, 
and, thus, only values of $K$ in the range $[-1,1]$ have to be considered. 

The ferromagnetic phase of the AT 
model breaks a $Z_4$ symmetry and is defined by $\langle \sigma \tau \rangle \neq 0$ and $\langle \sigma \rangle =  \pm \langle \tau \rangle$. The transition from this 
$Z_4$ ordered state to the fully disordered state ($\langle \sigma \tau \rangle = 0$ and $\langle \sigma \rangle =   \langle \tau \rangle = 0$) has continuously changing 
exponents which are exactly known as a function of $K$~\cite{Nienhuis, AT} using the following relations based on the powerful Coulomb-gas (CG) formulation for studying 
this class of 2D phase transitions (see Ref.~\onlinecite{Nienhuis} for an excellent review of this approach):
\be
y_t=2-2/g_R,~~~ y_h=15/8,~~~ y_p=2-1/(2g_R),
\label{cg}
\ee
with
\be
g_R=\frac{8}{\pi} \arcsin{\left (\frac{1}{2} \coth(2/T_c)\right )}
\ee
being the CG coupling. The critical temperature $T_c$ is exactly given by the fact that there is a self-dual line;
\be
\sinh(2/T_c)=\exp{(-2 K/T_c)}.
\ee
The three exponents are the thermal exponent $y_t=1/\nu$, the magnetic exponent $y_h=2-\eta/2$, which is fixed in the region, and the exponent
$y_p$ corresponding to a polarization field acting on one of the sets of Ising variables,  $P \sum_i \tau_i$ (which breaks the $Z_2$ symmetry 
of the Hamiltonian).\cite{Nienhuis} The corresponding scaling dimensions are $X_t=2-y_t$, $X_h=2-y_h$, $X_p=2-y_p$.

In the AT model, $K=0$ corresponds to the decoupled Ising limit and $K=1$ corresponds to $4$-state Potts model universality. When $K$ is extended from $0$ to negative 
values, the thermal exponent $\nu$ increases and the specific heat develops a cusp. In the AT model, the critical line is defined up to the point $K=-1$ where $\nu=2$. 
The transitions from $K=0$ to $K=-1$ can be viewed as a subset of the critical points in the anisotropic XY model discussed above. Since the exponent 
$\alpha/\nu \geq 0$ when $K \in [0,1]$, this suggests that if there are continuous phase transitions in the $J_1$-$J_2$ model, then these belong in the same 
universality class as critical points in the AT model in the range $K \in [0,1]$. Moreover, the specific heat exponent $\alpha/\nu$ seems to 
decrease smoothly as $g$ is increased~\cite{prl12} and its behavior with g indicates that the frustrated Ising model may have all the critical points of the AT model from 
$K=0$ till $K=1$ ($4$-state Potts model), which will then be the multicritical point in the frustrated Ising model and is the end point of the line of continuous 
transitions between $Z_4$ ordered and disordered phases in the AT model. 

As we will see later in Sec.~\ref{sec:MC}, the continuous transitions in the $J_1$-$J_2$ model for $g \in [g^*, \infty)$ indeed can be mapped to the transitions 
in the AT model when $K \in [1,0)$. We will asume that the stripe transition for $g < g^*$ is first order instead of some unlikely and alternative exotic behavior 
outside the known scenarios for $Z_4$ symmetry breaking. Results near $g = 1/2$ discussed in Section~\ref{sec:MC} suggest that the first-order transitions are 
very weak, and large system sizes may be needed to observe unambiguously the expected first-order scaling behaviors. In principle one cannot exclude, based
on the numerics alone, that there is some yet unknown type of continuous transition for $1/2 < g < g^*$.

\section{Cluster mean-field theory}
\label{sec:CMF}

\begin{figure}
\center{\includegraphics[width=4.25cm, clip]{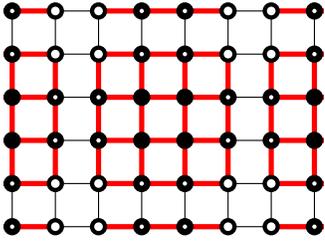}}
\vskip-1mm
\caption{(Color online) Illustration of a variational mean-field theory based on $4\times 4$ clusters. The infinite lattice is divided into clusters (sites connected
by the thicker, red lines), and a single isolated cluster with added magnetic fields (indicated here by different circles) is solved exactly (with no interactions 
between it and neighboring clusters), giving $\langle E^0_c\rangle_0$ and $F^0_c$. The energy $\langle E_c\rangle_0$ of the original system without the fields is 
evaluated using the mean-field decomposition $\langle \sigma_i\sigma_j\rangle_0$ $\to$ $\langle \sigma_i\rangle_0\langle\sigma_j\rangle_0$ for the bonds connecting 
clusters and using the imposed periodicity to translate both sites $i$ and $j$ into the same cluster. The fields are adjusted to minimize the upper bound 
$F^*_c=F^0_c+\langle E^0_c-E_c\rangle_0$ on the cluster free energy.}
\vskip-1mm
\label{embedded4}
\end{figure}

Here we study the $J_1$-$J_2$ model using a CMF approach based on $2\times 2$ and  $4\times 4$ clusters, as illustrated for
the latter case in Fig.~\ref{embedded4}.  In this section we will not just consider the striped phase and transition, but also the standard
ferromagnetic phase obtaining for $0 \le g \le 1/2$.

\subsection{Variational approach with a reference system}

One way to formulate a mean-field theory is to construct an approximate expression for the partition function with the aid of some solvable model.\cite{ChaikinLubensky} 
Let $E^0_\sigma$ be the energy for such a reference system in spin configuration $\sigma$. It is assumed that its partition function,
\begin{equation}
Z_0=\sum_{\sigma} {\rm e}^{-E^0_\sigma/T},
\end{equation}
can be calculated exactly in some way (numerically or analytically). We can write the partition function of the original system of
interest, with energy function $E_\sigma$, as
\begin{eqnarray}
Z &=& Z_0\sum_{\sigma} \frac{{\rm e}^{-E^0_\sigma/T}}{Z_0}{\rm e}^{-(E_\sigma-E^0_\sigma)/T} \nonumber \\
  &=& Z_0\langle {\rm e}^{-(E-E^0)/T}\rangle_0,
\label{zvarz0}
\end{eqnarray}
where $\langle \rangle_0$ denotes the expectation value with respect to the Boltzmann distribution of the model $E^0_\sigma$. Under our assumption,
this expectation value can also be evaluated exactly. A well known result is that \cite{ChaikinLubensky}
\begin{equation}
Z \ge Z_0{\rm e}^{-\langle E-E^0\rangle_0/T},
\end{equation}
from which we obtain an upper bound $F^*$ to the free energy $F=-T\ln(Z)$;
\begin{equation}
F \le F^* = F_0 +\langle E- E^0\rangle_0.
\label{fbound}
\end{equation}
This variational principle for the free energy is very useful if the reference model $E^0_\sigma$ has some parameters that can be varied. One can then 
minimize the free-energy bound $F^*$ with respect to those parameters, to obtain the best (in the sense of minimum free energy) variational solution 
to the system $E_\sigma$.

In the variational CMF approach the reference system is a infinite system is divided into clusters, with no interactions between the clusters. The energy 
of the infinite reference system can be written as a sum over the identical clusters $c$;
\begin{equation}
E^0=\sum_c E^0_c.
\end{equation}
A small isolated cluster can be solved by exact summation over all its spin configurations. The aim is to minimize the free energy of the $J_1$-$J_2$
model with respect to variational parameters of the reference system. In principle, the reference model can contain arbitrary field and spin dependent 
terms within the cluster. However, in practice, the cluster Hamiltonian function minimizing the variational free energy has exactly the same couplings as 
the original $J_1$-$J_2$ model, and only fields $-h_i\sigma_i$ acting on the edge spins are added. Since the $J_1$-$J_2$ model includes only up to
second-neighbor couplings, the edge here has the standard meaning of only the outermost layer of sites, but for longer-range interactions the
``edge'' extends further into the cluster.

Let us first consider independent fields $h_i$ coupling to all the spins $\sigma_i$, $i=1,\ldots,n$ within a cluster of $n$ sites (disregarding 
symmetries that will eventually imply that some of the fields should be equal);
\begin{equation}
E^0_c=\sum_{(i,j)}J_{ij}\sigma_{i}\sigma_{j}-\sum_{i=1}^nh_i\sigma_i,
\end{equation}
where $(i,j)$ refers to site pairs (counted only once) within the cluster. In the model we will consider explicitly, $J_{ij}=-J_1$ or $J_2$, but in 
principle $J_{ij}$ could include even longer-range interactions. The cluster energy defines the reference Boltzmann distribution with relative probabilities 
$W_0(\sigma)= {\rm exp}[-E^0_c(\sigma)/T]$ for the $2^n$ spin configurations $\sigma=\sigma_1,\ldots,\sigma_n$. Using this probability distribution, we can 
evaluate the partition function $Z^0_c=\sum_{\sigma}W_0(\sigma)$, $\langle E_0\rangle_0$, and the expectation value $\langle E\rangle_0$ of the original 
energy for each cluster. We have
\begin{equation}
\langle E_c\rangle_0 = \sum_{(i,j)}J_{ij}\langle \sigma_{i}\sigma_{j}\rangle_0 
+ \frac{1}{2}\sum_{(i,j)'}J_{ij}\langle \sigma_{i}\rangle_0\langle\sigma_{j}\rangle_0,
\end{equation}
where $(i,j)'$ in the second sum refers to interactions between a site $i$ in the cluster $c$ and a site in a different cluster. Since all clusters 
are equivalent, this site can be translated into an equivalent site $j$ of the cluster $c$. The factor $1/2$ accounts for the fact that each 
interaction bond $(i,j)'$ is shared by two different clusters. 

In the free-energy bound (\ref{fbound}) $F^*_c=-T\ln(Z^0_c)+ \langle E-E^0\rangle_0$ we need only the difference between the two energies, for which the 
intra-cluster interactions cancel;
\begin{equation}
\langle E-E^0\rangle_0 = \frac{1}{2}\sum_{\langle i,j\rangle'}J_{ij}\langle \sigma_{i}\rangle_0\langle\sigma_{j}\rangle_0 +\sum_{i}h_i\langle\sigma_i\rangle_0.
\end{equation}
To minimize $F^*_c$ we need its derivatives with respect to the fields;
\begin{eqnarray}
&&  T\frac{\partial F^*}{\partial h_k} = 
\sum_i h_i \Bigl ( \langle \sigma_i\sigma_k\rangle_0 -  \langle \sigma_i\rangle_0\langle \sigma_k\rangle_0 \Bigr )\nonumber \\
&&~~~~~~~~~~~~\sum_{\langle i,j\rangle'} J_{i,j} 
   \langle \sigma_i\rangle_0\Bigl ( \langle \sigma_j\sigma_k\rangle_0 -  \langle \sigma_j\rangle_0\langle \sigma_k\rangle_0 \Bigr ),
\label{fstarcderiv}
\end{eqnarray}
which can be written in the form
\begin{equation}
\frac{\partial F^*}{\partial h_k} = \sum_i \Bigl (h_i + \sum_{(j)_i} J_{ij}\langle \sigma_{j}\rangle_0 \Bigr )a_{ik} =0,
\end{equation}
where the notation $(j)_i$ in the second sum sum indicates summation for given $i$ over only those spins $j$ corresponding to inter-cluster (edge) 
interactions and $a_{ik}= \langle \sigma_i\sigma_k\rangle_0 -  \langle \sigma_i\rangle_0\langle \sigma_k\rangle_0$. These equations are satisfied if
\begin{equation}
h_i = -\sum_{(j)_i} J_{ij} \langle \sigma_{j}\rangle_0,
\label{hselfc}
\end{equation}
which amounts to self-consistency conditions for all the fields. For sites $i$ that have no non-zero inter-cluster interaction $J_{ij}$, we have $h_i=0$, 
i.e., we need to consider only fields on the edge spins. This self-consistent solution has the lowest free energy also if other interactions are allowed 
within the reference system $E^0$. The variational approach is therefore equivalent to the self-consistent approach. The advantage of starting from the 
variational ansatz is that the free energy (its upper bound $F^*_c$) is also obtained without any further assumptions.

\begin{figure}
\center{\includegraphics[width=7.5cm, clip]{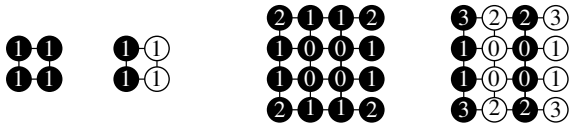}}
\caption{Clusters and fields used in mean-field calculations for the $J_1$-$J_2$ Ising model. For a ferromagnetic state (all black 
circles), all fields are positive, of strength $h_1$ for the $2\times 2$ cluster and $h_1,h_2$ for the $4\times 4$ cluster (with $h_0=0$, because the 
sites marked $0$ are not on the cluster edge). For the striped state, the black and white circles indicate positive and negative fields of magnitude 
$h_s$ and sign $(-1)^{x_i}$, where $x_i$ is the $x$-coordinate of the sites $i$ to which it couples. For the $4\times 4$ cluster, the broken rotational 
symmetry of the striped state implies one more variable field than for the ferromagnet, for a total of three adjustable parameters $h_1,h_2,h_3$.}
\label{mfvclusters}
\end{figure}

In practice one does not have to treat all the fields $h_i$ as independent parameters, because the optimal fields for an ordered state will obey symmetries 
corresponding to those of the order parameter. For the $J_1$-$J_2$ Ising model considered here, we have ferromagnetic and striped order for $J_2/J_1<1/2$
and $>1/2$, respectively. The field arrangements appropriate for these order parameters on $2\times 2$ and $4\times 4$ clusters are illustrated in 
Fig.~\ref{mfvclusters}. We will not consider the $3\times 3$ cluster, because it is not appropriate for the striped state (due to its incompatibility with the 
periodicity $2$ in one of the directions, although in principle one could also take this into account by modified boundary conditions). Using the appropriate 
symmetries, the largest number of parameters is here three, for the stripe order on the $4\times 4$ lattice. With such a small number of parameters, we can 
easily solve the self-consistency equations numerically.

\begin{figure}
\center{\includegraphics[width=8cm, clip]{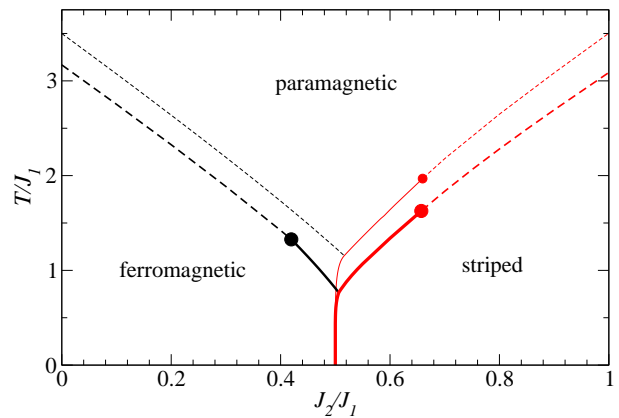}}
\caption{(Color online) Phase diagram of the $J_1$-$J_2$ Ising model in the coupling-temperature plane based on mean-field calculations with clusters of 
size $2\times 2$ (thin curves, at higher $T$) and $4\times 4$ (thick curves, at lower $T$). Dashed and solid curves indicate continuous and first-order phase 
transitions, respectively, with the circles indicating the multi-critical points at which the order of the transition changes.}
\label{j1j2phases}
\end{figure}

\subsection{Phase diagram}

By finding the optimal solutions for both ferromagnetic and striped field patterns, and comparing their free-energy minimums for a range of temperatures
and coupling rations $g$, the phase diagram of the system can be extracted. To precisely determine a phase boundary as a function of temperature at 
fixed $g$, one can carry out a bracketing procedure to locate the point at which the optimal solution changes between paramagnetic and ordered, 
or between ferromagnetic and striped ordered. Fig.~\ref{j1j2phases} shows the phase diagram obtained this way, based on both $2\times 2$ and $4\times 4$ 
clusters. There are continuous as well as first-order transitions. First-order transitions obtain close to $g=1/2$, which is the point at which we 
already concluded that there should be such a transition as a function of $g$ when $T\to 0$. The point at which the transition becomes continuous
is stable with respect to the cluster size, $g^* \approx 0.66$, and is in remarkably good agreement with the value $g^* \approx 0.67$ obtained in the 
previous MC work \cite{prl12} identifying the Potts point. The mean-field calculation can of course not give any information on the true critical exponents.

The paramagnetic--ferromagnetic transition is seen to always be continuous within the $2\times 2$ cluster calculations, but it also changes to 
first-order in a narrow window of $g$ values when the $4\times 4$ cluster is used. There are no clear indications from previous MC simulations
of the transition being first-order in this regime, but it may be worth examining this issue more carefully as well with improved MC simulations. 
In this paper we focus on the stripe phase for $g > 1/2$, however.

In principle one can try to extrapolate the critical temperature to infinite cluster size, but this is not possible based on just the $2\times 2$ 
and $4\times 4$ results. In principle one can add standard single-site MF calculations and the $3\times 3$ cluster to enable some estimates. Here,
however, our interest is in the order of the transitions and we will not attempt any CMF-based extrapolations of $T_c$ (which anyway come out very precisely
in the MC and TM calculations, as discussed in the later sections).

 \begin{figure}
\center{\includegraphics[width=8.5cm, clip]{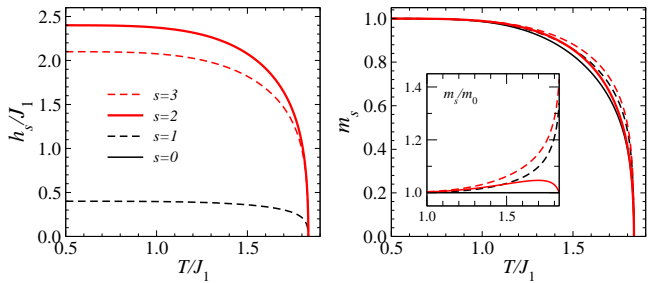}}
\caption{(Color online) Self-consistent variational field parameters (left) and the corresponding induced stripe magnetizations (right) 
versus temperature obtained using a $4\times 4$ cluster at $J_2/J_1=0.7$. The inset of the right panel shows the ratio of the shell magnetizations
for shells $s>0$ to the one with $s=0$ (center of the cluster). The shell index $s$ follows the convention illustrated in Fig.~\ref{mfvclusters}.}
\label{l4j70mh}
\end{figure}

\subsection{Order parameter and free energy}

An example of self-consistent field parameters and induced shell magnetizations is shown in Fig.~\ref{l4j70mh}. These results are 
for the $4\times 4$ cluster at $g=0.7$, where there is a continuous paramagnetic--striped transition at $T/J_1\approx 1.83$. All the fields vanish 
continuously at this point. Note that the order parameter is not uniform, as it should be for an infinite system, but shows significant variations between
the shells. For most of the temperature range, the order is the weakest at the four central spins, where there is no field, but close to $T_c$ one of the edge 
magnetizations becomes equal to it. The ratio of the magnetizations $m_s$ to the central magnetization $m_0$ is significantly different from $1$ 
close to $T_c$, but should remain finite because all $m_s$ should vanish as $T\to T_c$ with the mean-field power-law behavior $m_s \sim (T_c-T)^{1/2}$. In 
general, one would expect that the four central spins should most closely represent the behavior of the infinite system. If we could go to much larger cluster 
sizes, we would expect the order parameter to become uniform in the interior of the cluster, with non-uniformity emerging gradually as the edges are
approached.

\begin{figure}
\center{\includegraphics[width=8.4cm, clip]{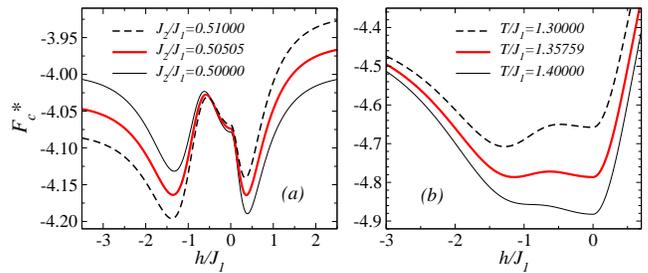}}
\caption{(Color online) Free energy versus the external field parameter $h$ in the $2\times 2$ cluster mean-field theory at and close to first-order
transitions. Positive and negative $h$ correspond to ferromagnetic and striped field patterns, respectively. (a) shows results for
different coupling ratios at and close to the ferromagnetic--striped transition at fixed $T/J_1=1$, and (b) shows the 
behavior for temperatures at and close to the striped--paramagnetic transition at $g=0.55$.}
\label{l2fcurves}
\end{figure}

To discuss the first-order transitions, it is useful to examine the free energy of the $2\times 2$ cluster, where there is just one variational
parameter. The free energy for both the ferromagnetic and striped fields can can be showed in the same graph by defining a new parameter $h$, such that 
for $h>0$ this is the ferromagnetic field $h_1=h$, whereas for $h<0$ it is the strength of the stripe field; $h_1=|h|$. Two examples of the dependence
of the cluster free energy $F^*_c$ on this parameter as a first-order transition is crossed are shown in Fig.~\ref{l2fcurves}. In the left panel, 
two minimums for $h\not=0$ can bee seen, corresponding to ferromagnetic and stripe orders, whereas in the right panel one of the minimums is at $h=0$, 
corresponding to the paramagnetic phase, and the other minimum is for a striped state. In either case, the two minimums are degenerate at the transition 
between the two phases. Changing $g$ at fixed $T$ (as in the left panel) or varying $T$ at fixed $g$ (right panel), the degeneracy is broken and one 
of the states becomes the stable one. The other, higher minimum then corresponds to a meta-stable state. 

\begin{figure}
\center{\includegraphics[width=8.4cm, clip]{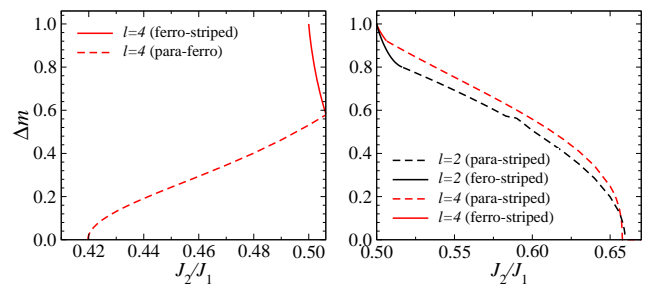}}
\caption{(Color online) Dependence on the coupling ratio of the discontinuity of the ferromagnetic magnetization (left) and stripe magnetization (right) 
at the first-order transitions obtained with the $2\times 2$ and $4 \times 4$ clusters ($l=2,4$). Note in the left panel and in Fig.~\ref{j1j2phases} that for 
$g$ in the range $0.50 \sim 0.504$, as $T$ is lowered there is afirst a paramagnetic--ferromagnetic transition, followed by a transition into the
low-temperature stripe state.}
\label{j1j2jumps}
\end{figure}

The discontinuities associated with the first-order transitions vanish continuously at the special multi-critical points indicated with circles in
the phase diagram in Fig.~\ref{j1j2phases}. Fig.~\ref{j1j2jumps} shows the behavior of all the discontinuities for the $2\times 2$ and $4\times 4$ clusters.

\section{Monte Carlo simulations}
\label{sec:MC}

We have simulated the $J_1$-$J_2$ Ising model using a standard single-spin Metropolis algorithm.~\cite{metroalg} Due to the presence of frustration, cluster MC
methods cannot be used for this model unless $J_1 (J_2) =0$. We found that single-spin Metropolis algorithm is sufficient to study the thermal phase transitions accurately 
if $g$ is not very close to $1/2$ (we have gone up to $g=0.52$ using single spin-flip MC moves). The transitions closer to $g=1/2$ can be simulated using a combination
of parallel tempering and certain non-local spin flips~\cite{honecker1} which have a high acceptance probability very close to $g=1/2$. We have also simulated the AT model 
on the square lattice, and for that we again use a local Metropolis algorithm except at $K=1$, where we use a cluster algorithm.~\cite{SW} Temperature is measured in units 
of $J_1$ for the frustrated Ising model and in units of $K$ for the AT model. MC simulations combined with finite-size scaling and universality arguments provides 
the most unbiased method to understand the nature of the transitions in the $J_1$-$J_2$ model. 

\begin{figure}
\center{\includegraphics[width=8.4cm, clip]{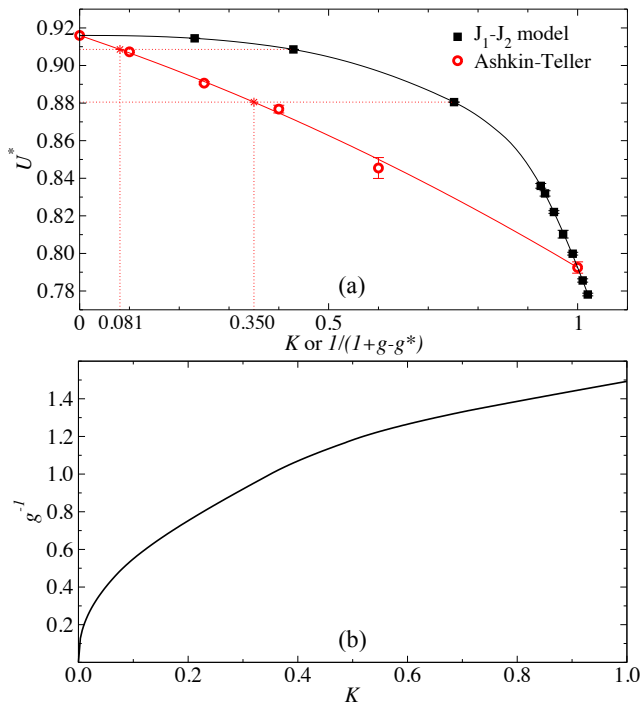}}
\vskip-1mm
\caption{(Color online) (a) Universal crossing value $U^*$ of the Binder cumulant vs $K$ for the AT model and vs $1/(1+g-g^*)$, $g^*=0.67$ for the 
$J_1$-$J_2$ model. The AT data points have been fitted to a single polynomial function, while the $J_1$-$J_2$ data were fitted with several polynomials 
in segments. The corresponding $K$ values with the same $U^*$ as $g=1.0$ and $g=2.0$ are marked in the graph. The map between the two 
models at these points are: ($g=1.0$) $1/(1+g-g^*)=0.752$ to $K=0.35$ with $U^*=0.8805$; ($g=2.0$) $1/(1+g-g^*)=0.429$ to $K=0.081$ with $U^*=0.9086$. (b) 
Map between the AT model ($K$) and the $J_1$-$J_2$ model ($g^{-1}$) using the universality of $U^*$ and the procedure illustrated in (a).}
\label{map2}
\vskip-3mm
\end{figure}

\subsection{Calculated Observables}

Before proceeding further, we define the observables that we measure in our MC simulations for the $J_1$-$J_2$ and the AT models respectively. Let us first
define the order parameters that characterize the broken $Z_4$ phase in both the models. The striped phase of the 2D frustrated Ising model is characterized by 
a two-component order parameter $(m_x,m_y)$ with 
\be
m_x &=& \frac{1}{N} \sum_{i=1}^N \sigma_i (-1)^{x_i}, \\
m_y &=& \frac{1}{N} \sum_{i=1}^N \sigma_i (-1)^{y_i},
\ee
where $(x_i,y_i)$ are the coordinates of site $i$ on a $L \times L$ periodic square lattice and $N=L^2$. We define $m^2=m_x^2+m_y^2$ and the stripe susceptibility as 
\be
\chi = \frac{N}{T}(\langle m^2 \rangle -\langle|m| \rangle^2).
\ee
We also measure the specific heat, 
\be
C_v = \frac{N}{T^2}(\langle E^2 \rangle -\langle E \rangle^2),
\ee
where $E$ is the energy per site. For the AT model, the order parameter can again be expressed as a 2D vector $(m_{\sigma},m_{\tau})$ where 
\be
m_{\sigma} &=& \frac{1}{N} \sum_{i=1}^N \sigma_i, \\
m_{\tau} &=& \frac{1}{N} \sum_{i=1}^N \tau_i,
\ee
and $m^2=m_{\sigma}^2+m_{\tau}^2$. The order parameter susceptibility $\chi$ and specific heat $C_v$ are then defined exactly in the same way as described
above for the $J_1$-$J_2$ model. 

We also compute the Binder cumulant of the order parameter in  both models. It is defined as
\be
U = 2 \left(1-\frac{1}{2} \frac{\langle m^4 \rangle}{\langle m^2
    \rangle^2}\right), 
\label{binderdef}
\ee
where the constants are chosen to give a step-function ($U \to 0$ and $U \to 1$ in the disordered and ordered phase, respectively) for a 2D vector order parameter 
in the thermodynamic limit.\cite{Sandvik10} 

Lastly, we collect the histograms of the squared order parameter $m^2$ and the energy $E$ near the transition for both the models. These are
helpful for analyzing the pseudo-first-order behavior in detail.

\subsection{Map between AT and $J_1$-$J_2$ critical points}

The Binder cumulant, Eq.~(\ref{binderdef}), turns out to be especially useful in establishing the universality class of the continuous transitions in the $J_1$-$J_2$ model. 
It is well known~\cite{Goldenfeld} that for continuous phase transitions, the Binder cumulant for different system sizes cross at the critical point (when the system 
size is large enough). The value of the crossing $U^*$ is universal as well and characterizes the universality class of the phase transition. $U^*$ may in some cases depend 
on details of the model beyond the universality class, like the boundary conditions and shape of the lattice and the anisotropy of the interactions.~\cite{Ustar}
However, in our case, both the $J_1$-$J_2$ model and the AT model live on periodic square lattices and the interactions respect the full symmetry of the lattice, 
so a comparison of $U^*$ between the two models seems to be justified. We have established this directly from our MC data by using the equality of $U^*$ to map phase 
transitions in one model to the other and then directly looking at critical exponents to check if the universality class is indeed the same.

We estimate $U^*$ from our MC simulations by extracting the crossing point of $U$ between data for pairs $(L,2L)$ and then extrapolating to 
$L \rightarrow \infty$.\cite{prl12} In Fig.~\ref{map2} we show $U^*$ as a function of the coupling $K$ and $g$ for the AT model and the $J_1$-$J_2$ model, 
respectively. This immediately establishes a numerical map between the continuous transitions of both the models. From Fig.~\ref{map2}, we see that $g \approx 0.67$ 
corresponds to the $4$-state Potts model universality in the AT model ($K=1$). This was already reported in Ref.~\onlinecite{prl12} and other consistency checks 
were used there to show that the multicritical point is located at $g^* = 0.67 \pm 0.01$. Since then, the location of $g^*$ has also been computed in 
Ref.~\onlinecite{honecker3}, and the result agrees perfectly with the earlier result.
 
\begin{figure}
\center{\includegraphics[width=8.4cm, clip]{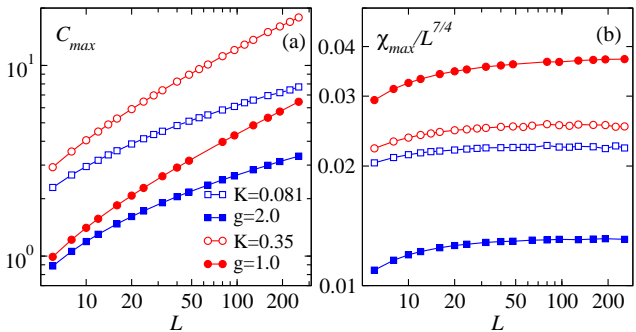}}
\vskip-1mm
\caption{(Color online) Examples of divergent peak-values of the specific heat and stripe susceptibility vs $L$ for the $J_1$-$J_2$ and AT models. The factor $L^{7/4}$
corresponding the asymptotic AT scaling of the susceptibility have been divided out in the right panel. Two point pairs are chosen for which the mapping in 
Fig.~\ref{map2} show that the critical exponents for the two models should coincide.} 
\label{compareKg}
\vskip-3mm
\end{figure}
 
As a further illustration of the correctness of the procedure, we use the data presented in Fig.~\ref{map2} to note that the phase transition at $g=1$ should map 
to $K \approx 0.35$ and $g=2$ to $K \approx 0.081$. This is indeed consistent with the divergent critical forms of physical quantities. In Fig.~\ref{compareKg} we 
plot the peak value of the specific heat $C_{\rm max}(L)$ and the order parameter susceptibility $\chi_{\rm max}(L)$ versus $L$ for the two models at the above points. By 
standard finite-size scaling arguments, $C_{\rm max} \sim L^{\alpha/\nu}$ and $\chi_{\rm max}(L) \sim L^{\gamma/\nu}$. For the system sizes studied here ($L \leq 256$), 
the exponent $\alpha/\nu$ for the two models, estimated from the slope of $C_{\rm max}(L)$ on a log-log scale (Fig.~\ref{compareKg}, left panel), converges to the same 
value in a very similar way for the $J_1$-$J_2$ and Potts model at the corresponding $g$ and $K$ values. This is also the case for the exponent $\gamma/\nu$ 
(Fig.~\ref{compareKg}, right panel), which converges to the value $7/4$, as is expected for AT universality. The latter behavior is of course less useful, 
since the exponent $\gamma=7/4$ is expected on the whole critical curve. In both cases some deviations from pure power-laws can be observed, and we will discuss this 
further below. The behavior seen for the specific heat is nevertheless quite telling and suggestive of the same critical exponent in the two models at the mapped 
points. Thus, we have rather convincingly established the map (Fig.~\ref{map2}) between the parameters of the AT model and the frustrated $J_1$-$J_2$ Ising model 
in two different but mutually consistent ways. Note that this kind of map does not imply microscopic equivalence of the two models, which holds only in the weakly 
coupled Ising limit ($1/g \to 0$), but demonstrate common low-energy descriptions of the systems for the mapped parameter values.

\subsection{$4$-state Potts scaling at $g^*$}

\begin{figure}
\center{\includegraphics[width=8.4cm, clip]{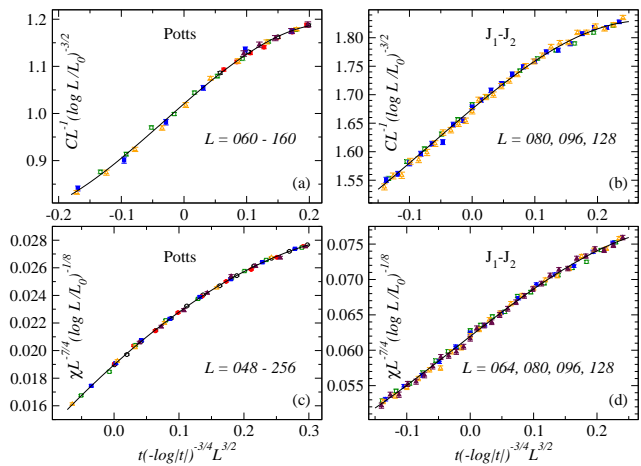}}
\vskip-1mm
\caption{(Color online) Data collapse with the anticipated leading logarithmic correction to the specific heat $C$ and the susceptibility $\chi$ for the $4$-state Potts 
model and the $J_1$-$J_2$ model at $g=0.68$. The system sizes included in (a) are  $L=60$, $64$, $80$, $120$, $128$, $160$, and in (c) $L=48$, $64$, $96$, $128$, 
$192$, $256$, while for (b) and (d) all the sizes are listed in the panels. The black curves are common curves fitted to all data points shown in each figure. The 
corresponding reduced chi-square per degree of freedom for the fitted curves in (a), (b), (c) and (d) are $\chi^2 = 1.3$, $1.6$, $1.2$, $1.8$. } 
\label{collapse}
\vskip-3mm
\end{figure}

To further strengthen the case for $g^* = 0.67 \pm 0.01$ being in the $4$-state Potts universality class, we next consider scaling in the form of data collapse of 
the specific heat and the susceptibility at the coupling $g=0.68$ of the $J_1$-$J_2$ model (for which we have MC data; the estimate for $g^*$ is based on interpolation,
as shown in Fig.~\ref{map2}) and for the $4$-state Potts model on the square lattice ($K=1$ in the AT model). The critical exponents of the $4$-state Potts model 
are:\cite{Salas} $\nu={2}/{3}, {\alpha}/{\nu} = 1, {\gamma}/{\nu} = {7}/{4}$. There are however important multiplicative logarithmic scaling corrections at this 
critical point that strongly affect finite-size scaling. According to Ref.~\onlinecite{Salas}, the divergences in the thermodynamic limit are of the forms:
\begin{eqnarray}
\xi &\sim& |t|^{-2/3} (-\log|t|)^{1/2}  \\
C_v &\sim& \frac{\xi}{(\log \xi)^{3/2}}, \\
 \chi &\sim& \frac{\xi^{7/4}}{(\log \xi)^{1/8}},
\label{logs}
\end{eqnarray}
where $t=(T-T_c)/T_c$  is the reduced temperature and $\xi$ is the correlation length. Then, using finite-size scaling arguments, one expects 
$C_v L^{-1} (\log (L/L_0))^{-3/2}$ and $\chi L^{-7/4}(\log(L/L_0))^{1/8}$ to be functions of the argument $t(-\log|t|)^{-3/4}L^{3/2}$. The two quantities with these expected asymptotic $L$-dependences divided out are graphed 
versus $t(-\log|t|)^{-3/4}L^{3/2}$ for different system sizes $L$ in Fig.~\ref{collapse}, with the non-universal scale factor  $L_0$ (which can be different for 
different quantities) treated as a fitting parameter to optimize the data collapse.

Fig.~\ref{collapse}(a) shows the resulting data collapse of $C_v$ for the $4$-state Potts model with a set of moderate to large system sizes $L=60 - 160$ included in the 
fitting procedure (with smaller sizes excluded because they are visibly affected by subleading corrections to scaling). The data collapse to a common fitted polynomial is 
statistically sound with the parameter $L_0=0.20 \pm 0.01$. Fig.~\ref{collapse}(b) shows the same kind of analysis for the $J_1$-$J_2$ model at $g=0.68$. The system sizes 
included here are in the range $L=80 - 128$ and $L_0=0.144 \pm 0.006$. The data collapse of $C_v$ in both the cases, using the same expected exponents and multiplicative 
logarithmic corrections, is another strong indication that $g=0.68$ in the $J_1$-$J_2$ model is in close neighborhood of the $4$-state Potts end-point of the critical 
Potts--Ising line of the AT model. 

The logarithmic scaling correction for the susceptibility, Eq.~(\ref{logs}), does not yield a good data collapse for either $g=0.68$ or the $4$-state Potts model. 
Therefore, instead of using $\chi L^{-7/4}(\log L/L_0)^{1/8}$ on the $y$-axis, we treat the exponent of the logarithmic function as another variable $r$ in addition to 
$L_0$. After carrying out a multi-variable data collapse, we obtain $r$ close to $-1/8$, instead of the value $1/8$ proposed in Ref.~\onlinecite{Salas}. 
Figs.~\ref{collapse}(c), (d) show the results for the two models with $r = -1/8$. Given that the data collapse is very good in both cases, and only the sign
of the log-exponent differs from what was expected, the natural conclusion is that there is a sign mistake in the analytical result for this exponent in
in Ref.~\onlinecite{Salas}. Thus, we propose that $\chi \sim \xi^{7/4}(\log\xi)^{1/8}$, instead of the form in Eq.~(\ref{logs}).

\begin{figure}
\center{\includegraphics[width=8cm, clip]{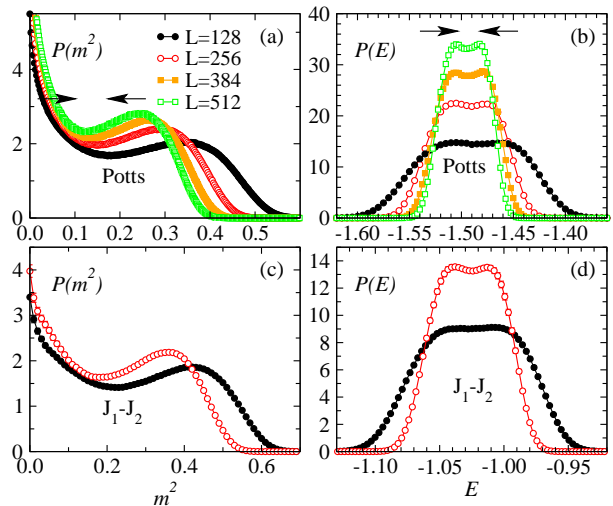}}
\vskip-1mm
\caption{(Color online) Histograms of the squared order parameter $m^2$ and the energy $E$ for (a), (b) the 4-state Potts model (the $K=1$ AT model) and (c),(d) for the 
$J_1$-$J_2$ model at $g=0.67$. Here $T$ is very close to $T_c$, chosen such that the two peaks in the energy histograms are of the same height; for the Potts model 
$T/K=3.64231$ for $L=128$ and $3.64460$ for $L=256$, while for the J1-J2 model $T/J1=1.2014$ for $L=128$ and $1.2004$ for $L=256$. }
\label{histo}
\vskip-3mm
\end{figure}

\subsection{ Pseudo-first-order behavior in the AT model} 

Even though the transitions in the AT model are all continuous, there are interesting pseudo-first-order signatures at finite system sizes at the $4$-state Potts point
and its neighborhood. This was explicitly shown in Ref.~\onlinecite{prl12} using various observables. Here, to further investigate this behavior, we show histograms of 
the distribution of the squared order parameter $m^2$ and energy per site $E$ for the $4$-state Potts model in the top panels of Fig.~\ref{histo}. For each system size, 
the temperature is very close to $T_c$, chosen to ensure that the two peaks in the energy histograms are of the same height. There is clearly a double-peak structure 
present even for the large system sizes $L=512$. The distance between the peaks decreases slowly as the system size increases and the dip between the peaks 
also does not increase appreciably (as it should if the double-peak structure is evolving into delta-functions corresponding to phase coexistence at a
first-order transition). This can in principle happen for a weak first order transitions as well, if system sizes used are $L \ll \xi$, where $\xi$ is the 
large but finite correlation length at the transition. However, in this case we know rigorously that the $4$-state Potts model harbors a continuous transition. For a 
continuous transition, there cannot be an order parameter jump or a latent heat in the thermodynamic limit. Thus, the distance between the double peaks will eventually 
shrink to zero when $L \rightarrow \infty$. This type of double peak structure was previously also observed in the energy distribution of the Baxter-Wu model on the 
triangular lattice,~\cite{baxterwu} which is in the same universality class as the $4$-state Potts model. The corresponding histograms of the $J_1$-$J_2$ model at 
$g = 0.67$ (which equals $g^*$ within error bars) also show a very similar behavior as can be seen from figures in the bottom panel of Fig.~\ref{histo}, again confirming
the equivalence of the critical behaviors with that of the $4$-state Potts model.

In Ref.~\onlinecite{prl12}, it was also shown that the Binder cumulant of the order parameter exhibits a non-monotonic behavior with $T$, developing a negative peak 
at $K=1.0$ and its vicinity (e.g., at $K=0.95$) in the AT model. A negative Binder peak is often taken as evidence of first-order transition, but here the transitions 
are clearly continuous. However, the negative peak increases very weakly with system size $L$, with the increase being much slower than the expected \cite{Vollmayr} 
$L^2$ divergence. Moreover, the dip is more pronounced at $K=1$ compared to $K=0.95$, which indicates that there may be a $K^*$ below which these pseudo-first-order 
signatures vanish. These pseudo first order signatures in turn lead to an over-estimation of the region of first-order transitions when the appearance of double-peak 
structures in energy or order-parameter histograms are taken to be indicative of discontinuous transitions in the $J_1$-$J_2$ model (see Ref.~\onlinecite{prl12} for 
more discussion on this point). 

Note that a two-peak structure was found in the energy histogram at very large system sizes for $g=0.9$ in Ref.~\onlinecite{honecker2}, 
which was taken as evidence of the first-order transition extending at least up to this value; based on our conclusions we instead see that the pseudo-first-order region 
in the $J_1$-$J_2$ model extends from $g^* \approx 0.67$ to $g \lesssim 1$. Following our previous work, Ref.~\onlinecite{honecker3} recently considered the energy 
histograms at $g=0.8$ for even larger system sizes than before. A double-peak structure appears when the system size reaches about $L=1000$, but going further it 
eventually disappears again, around $L=2000$. This again confirms the pseudo-first-order behavior the $J_1$-$J_2$ model close to $g^*$, and is a demonstration
of pitfalls in distinguishing between continuous and first-order transitions. We stress here that the reason we were able to avoid this pitfall is that the Potts 
model is rigorously known to harbor a continuous transition, and we found that the behavior of the $J_1$-$J_2$ model at $g=g^* \approx 0.67$ matches it very well in
all respects. Thus, the combination of analytical theory and numerics was crucial.

\begin{figure}
  \center{\includegraphics[width=8.4cm, clip]{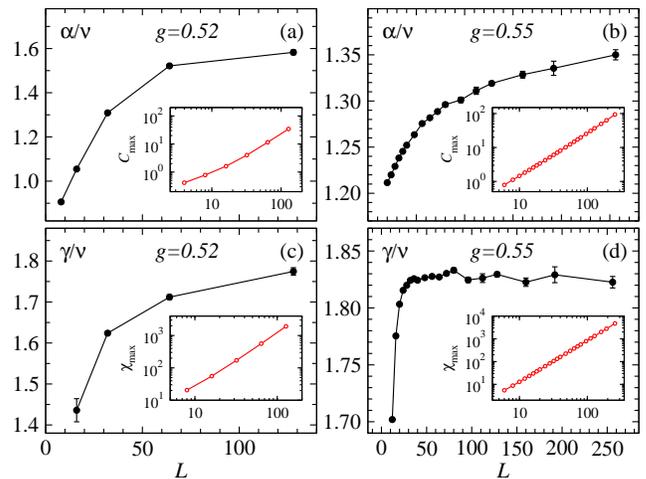}}
  \vskip-1mm
  \caption{(Color online) Scaling exponents $\alpha/\nu$  and $\gamma/\nu$ of $C_{\rm max}(L)$ and  $\chi_{\rm max}(L)$. The exponent 
    $\alpha/\nu$ is calculated based on the {\it local} slope of $C_{\rm max}(L)$ between $L$ and $L/2$; $\gamma/\nu$ is calculated similarly from
    $\chi_{\rm max}(L)$. The system sizes $L \le 128$ in (a) and (c); and $L \le 256$ in (b) and (d). }
  \label{localSlope}
  \vskip-3mm
\end{figure}

\subsection{Weak first-order transitions} 

Since the $J_1$-$J_2$ model at $g^*$ is in the $4$-state Potts universality class and approaches the Ising limit for higher $g$, the transitions for $1/2<g<g^*$ 
have to be first-order transitions if there is to be a correspondence with the known scenarios for $Z_4$-breaking transitions discussed in Sec.~\ref{sec:universality}. 
Unless some unknown scenario applies, which we find unlikely (but cannot rule out completely), 
all the transitions in the range $1/2 < g < g^*$ are very weak first-order transitions. Weakening of 
a discontinuous transition is expected when approaching a multicritical point, since the continuous transition has to be approached in a continuous manner. However, 
here the transitions in the close neighborhood of the obvious first order point $g=1/2$ are {\em also} weakly first order. Ref.~\onlinecite{honecker1} suggested this 
based on the appearance of double peak structure in the energy histograms. However, as we saw, such double-peak structure also appears at the Potts point and its 
neighborhood (where they disappear in the thermodynamic limit). Here we show the evolution of the {\em pseudo critical exponents} with system size $L$ on the (likely) 
first-order side close to the $g=1/2$ point.

We analyze the peak value $C_{\rm max}(L)$ of the specific heat and $\chi_{\rm max}(L)$ of the stripe susceptibility. By finite-size scaling arguments for first-order 
transitions, these quantities should diverge as $L^2$ in 2D.~\cite{Vollmayr} Examples of the scaling behavior are shown in Fig.~\ref{localSlope}. Two coupling ratios 
$g=0.52$ with system size $L \le 128$ and  $g=0.55$ with system size $L \le 256$ are considered. The peak value of $C_{\rm max}(L)$ and  $\chi_{\rm max}(L)$ are shown 
as the inset in each graph on a log-log scale. Graphed in this way, the peak value of $C_{\rm max}(L)$ and  $\chi_{\rm max}(L)$  seem to follow a linear scaling behavior, 
especially in the insets of Figs.~\ref{localSlope}(b), (c), (d). A more systematic analysis involves extracting the running exponents $\frac{\alpha}{\nu}(L)$ and 
$\frac{\gamma}{\nu}(L)$ from the {\it local} slope of $C_{\rm max}(L)$ and  $\chi_{\rm max}(L)$ between, e.g., system sizes $L$ and $L/2$. These should approach $2$ 
as $L \rightarrow \infty$ for a first-order transition. The first-order exponent $2$ is not obtained in these figures for the system sizes studied. The scaling exponents 
$\frac{\alpha}{\nu}(L)$  and $\frac{\gamma}{\nu}(L)$ [Figs.~\ref{localSlope}(a), (c)] increase as the system size increases for $g=0.52$, but have not converged at 
size $L=128$. This suggests that it may require very large system sizes to observe the expected $L^2$ scaling behavior. The scaling exponent $\frac{\alpha}{\nu}(L)$ 
for $g=0.55$ [Fig.~\ref{localSlope}(b)] is further away from $2$ at the same system size $L=128$, while it shows the same tendency to increase as $g=0.52$. The scaling 
exponent $\frac{\gamma}{\nu}(L)$ for $g=0.55$ actually seems to have in fact converged to $\approx 1.82$, at a comparably smaller system size $L=80$. This is again 
may be indicative of the large correlation length involved at a very weak first-order transitions, this being reminiscent of extremely weak first-order transitions like that
in the $5$-state Potts model~\cite{peczak} on the square lattice. Itt is puzzling, however, that there is no visible size dependence in the local $\frac{\gamma}{\nu}(L)$ 
between $L=80$ and $256$ for $g=0.55$. It would be interesting to go to larger lattices.

Note that for both couplings $\frac{\alpha}{\nu} (L)$ is already much larger than $1$, which is 
the maximum divergence expected if the critical point is continuous and in the AT universality class. It is an interesting open question to 
understand the mechanism which makes the transitions near $g=1/2$ so weakly first-order; it likely is related to the extensive degeneracy of the system at $T=0$.

\section{Transfer Matrix calculations}
\label{sec:TMF}

We now address the stripe transition in the $J_1$-$J_2$ model using numerical TM calculations. 
Consider the lattice wrapped on a cylinder of infinite length with circumference $L$, on which the TM is constructed. It is possible 
to perform finite-size scaling in $L$ to obtain properties of the phase transition like critical exponents. We use a sparse-matrix 
technique \cite{sparsetm1,sparsetm2}  to enable the calculations of sizes up to $L=26$. Note that using TM calculations one can 
in principle obtain the exact non-mean field exponents of the phase transition by using sufficiently large $L$. This is unlike the 
CMF approach that we discussed in Sec.~\ref{sec:CMF} where the exponents are mean field exponents (although for sufficiently large 
clusters the true exponents should emerge close to the transition, but such large clusters cannot be reached in practice).

The critical exponents of the transition are obtained by calculating three types of scaled gaps based on the eigenvalues of the TM. Each of them 
converges to a separate scaling dimension when system size $L$ tends to infinity at the critical point. The antiferromagnetic scaled gap is defined as
\begin{eqnarray}
X_h (T,g,L)=
{L \over 2 \pi}\mbox{ln}\left (\frac{\lambda_0}{\lambda_1}\right),
\end{eqnarray}
where $\lambda_0$ is the largest eigenvalue and $\lambda_1$ the largest eigenvalue in the subspace that breaks the symmetry of two neighboring sites,
(i.e., corresponding to the stripe state), which means that the associated eigenvector $\vec{v}_1$ satisfies
\begin{eqnarray}
\vec{v}_1=- {\mathbf R} \vec{v}_1,
\end{eqnarray}
where ${\mathbf R}$ is the translation operator that translates the lattice by one unit along the axis of the cylinder. Thus, the scaled gap is also 
called the {\it stripe scaled gap}. For the system to host this stripe order, we have to restrict the system to even $L$. In addition, the eigenvector 
also bears odd parity when the system is reflected about the center, but is invariant under global spin flips.

The two other scaled gaps are defined as
\be
X_{t_1}(T,g,L)=\frac{L}{2\pi} \ln \left ({\frac{\lambda_0}{\lambda_2}}\right ),
\ee
and 
\be
X_{t_2}(T,g,L)=\frac{L}{2\pi} \ln \left ( {\frac{\lambda_0}{\lambda_3}} \right ),
\ee
where $\lambda_2 $ and $\lambda_3$ are the leading and sub-leading eigenvalues associated to eigenvectors that are invariant under the lattice 
translation and global spin flips. However, it is not {\it a priori} clear which of these gaps corresponds to the thermal scaling dimension, and 
what the physical meaning is of the other gap.

According to finite-size scaling theory\cite{fss} and conformal invariance,\cite{conformal} the gap $X_i(T, g, L)$ in the vicinity of a 
critical point scales as
\be
X_i(T,g, L)=X_i+ a (T-T_c) L^{y_t}+ b u L^ {y_u} +\cdots,
\ee
where $i$ indicates one of the three gaps ($i=h, i=t_1$, or $i=t_2$),  $y_t$ is the leading thermal exponent, $u$ the leading irrelevant field, 
and $y_u$ is the associated irrelevant exponent. The constants  $a, b$ are unknown (not universal).

We calculate the scaled gap $X_h(T, g, L)$ and then numerically solve for $T_c(L)$ using the following scaling equation:
\be
X_h(T,g,L)=X_h(T,g,L-2).
\label{scleq}
\ee
The solution $T_c(L)$ converges to the critical point $T_c$ as $L \to \infty$ in the following way:
\be
T_c(L)=T_c + a' u L^{y_u-y_t}+\cdots,
\ee 
where $a'$ is an unknown constant. We thus determined the critical points of the $J_1$-$J_2$ model for various $g$ values to good accuracy, and the
results are listed in Table \ref{j1j2tab}. These $T_c$ values extracted from the TM approach agree very well with our MC results.  

The scaled gaps $X_h, X_{t_1}$ and $X_{t_2}$ at the solutions $T_c(L)$ are calculated for a sequence of systems up to $L=26$. Generally speaking, these gaps 
should converge to the corresponding scaling dimensions, respectively, in the following way:
\be
X_i(L)=X_i+b' L^{y_u} +\cdots,
\label{xscl}
\ee
where $b'$ is an unknown constant. However, at the 4-state Potts point, the irrelevant exponent $y_u$ is zero, i.e., the corresponding field is 
marginally irrelevant, which leads to the following multiplicative logarithmic correction to scaling:\cite{Salas}
\be
X_i(L)=X_i+ {b_1' \over \ln L} + {b_2' \ln(\ln L) \over (\ln L)^2}
+\cdots,
\label{logcorr}
\ee
with $b_1', b_2'$ unknown constants.

From the scaling analysis of our MC data (Sec. ~\ref{sec:MC}), we already know that $g^* \approx 0.67$. Fitting $X_i(L)$ according to Eq. (\ref{xscl}) 
for $g> 0.67$, we obtain the scaling dimensions $X_h, X_{t_1}, X_{t_2}$.  The convergences of our data is not very good. This is because the irrelevant exponent 
$y_u$ has a small absolute value even away from $g^*$. This is also the case for the AT model when $K$ is close to $1$ (the 4-state Potts critical point). For $g=0.67$, 
the scaling dimensions are estimated by fitting $X_i(L)$ to Eq. (\ref{logcorr}). The results of such fits are listed in Table \ref{j1j2tab}.

\begin{figure}
\includegraphics[width=8.4cm, clip]{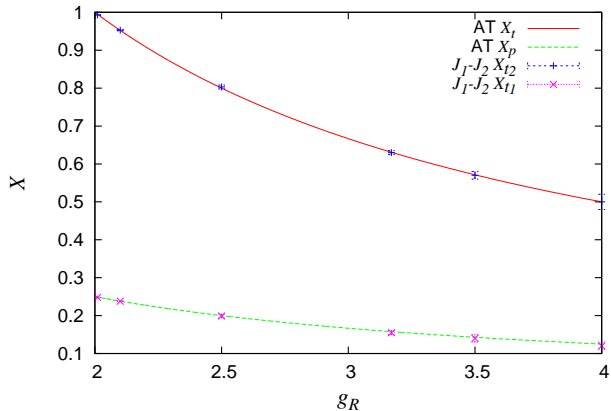}
\caption{ (Color online) The scaling dimensions as functions of the Coulomb gas 
coupling $g_R$. The curves for the AT model are theoretical predictions, while points are numerical results 
for the $J_1$-$J_2$ model with numerical TM-estimated $g_R$.}
\label{xg}
\end{figure}

It is remarkable that, for all $g$ in the region $[0.67, 5]$, the ratio of 
$X_{t_2}$ and $X_{t_1}$ is always close to 4. 
Comparing with the CG formula Eq. (\ref{cg}) describing the AT 
universality class, we thus identify $X_{t_2}$ as the thermal scaling 
dimension, and $X_{t_1}$ a scaling dimension corresponding to the polarization 
scaling dimension $X_p$ of the AT model. 
Meanwhile the striped scaling dimensions are close to 1/8 for all $g$, which
corresponds to the magnetic scaling dimension of the AT model.
We further obtain $g_R$ for each $g$ using Eq. (\ref{cg}) which
expresses $X_p$ as a function of the CG coupling $g_R$. 
These results are also listed in Table \ref{j1j2tab}. 
We plot our numerical results $X_{t_1}, X_{t_2}$ versus $g_R$ in 
Fig. \ref{xg}, together with the CG predictions of $X_t$ and $X_p$ for
the AT model.  Thus, the TM calculations give a  picture consistent with the
MC simulations when we consider the ratio of the scaling dimensions
$X_{t_2}/X_{t_1}$. The TM calculations, however, converge very slowly with $L$ near the 4-state
Potts point, which makes the method  unsuitable for the extraction of
$g^*$ itself. 

\section{Conclusions}
\label{sec:summary}

We have shown that the thermal transitions from the striped ordered phase in
the $J_1$-$J_2$ model for the range $g \in [g^*,\infty)$ can be fully mapped
to the continuous phase transitions of the well-known AT model. The special point 
$g^* \approx 0.67$
 corresponds to the 4-state Potts universality class and for $g \rightarrow \infty$ 
the transition approaches the standard Ising universality class. We have provided a numerical mapping
between the critical lines of the two models, based on matching universal properties; critical
exponents as well as order-parameter histograms. 

Interestingly, the 4-state Potts model and the 
neighboring transitions in the AT model show a pseudo-first-order behavior on finite lattices, 
though these transitions are rigorously known to be continuous. The energy and order-parameter 
histograms show double-peak structures near $T_c$, with the distance between the peaks decreasing
slowly to zero as the system size is increased. This feature of the Potts point and its neighborhood 
consequently leads to similar effects in the $J_1$-$J_2$ model as well, in the vicinity of $g^*$. 
This feature was misinterpreted as indicative of first-order transitions in some previous studies. 
The frustrated Ising model exhibits this pseudo-first-order behavior for $g^* \leq g \lesssim 1$. 

\begin{table}
\caption{Best estimates for the critical properties obtained using the TM method for the $J_1$-$J_2$ model 
at various $g$ values.}
\begin{tabular}{l|l|l|l|l|l}
\hline
\hline
$g$    &$|J_1|/T_c$& $X_h=\eta/2$ & $X_{t_2}=2-1/\nu$  &$X_{t_1}$  & $g_R$\\
\hline
0.67   &0.8335(2)  & 0.12(1)     & 0.50(2)        & 0.12(1)   & 4     \\
0.70   &0.7758(1)  & 0.12(1)     & 0.57(1)        & 0.14(1)   & 3.5   \\
0.75   &0.69866(5) & 0.12(1)     & 0.630(5)       & 0.155(5)  & 3.17  \\
1.00   &0.48029(5) & 0.123(5)    & 0.803(5)       & 0.199(5)  & 2.5  \\
2.00   &0.22468(3) & 0.125(2)    & 0.953(2)       & 0.238(1)  & 2.1  \\
5.00   &0.088406(5)& 0.125(2)    & 0.993(1)       & 0.248(1)  & 2.01 \\
\hline
\end{tabular}
\label{j1j2tab}
\end{table}

We further showed that the MC data of the $J_1$-$J_2$ model at $g^*$ can be scale-collapsed 
by using the critical exponents of the 4-state Potts model, if logarithmic scaling corrections
known to exist at this point are properly taken into account. CMF and TM calculations 
were also used to understand aspects of the phase transition. CMF on $2 \times 2$ and $4 \times 4$ 
clusters predict a multicritical point at $g^* \approx 0.66$; very close to the exact location based
on the MC calculations. However, since these calculations are mean-field in nature, the universality 
class of the continuous exponents cannot be determined directly with this approach. The TM calculations,
carried out on cylinders of infinite length and finite width, lead to accurate (well converged) results 
for the transition temperature $T_c$ and also present an alternative method (to MC simulations) to 
calculate critical exponents when $g$ is not too close to $g^*$. However, it is difficult to reliably 
compute the location of $g^*$ with the TM method based on accessible cylinder widths, due to effects 
of the logarithmic corrections discussed above.

An open  issue requiring further investigation is to understand why the transitions in the whole 
region $(1/2,g^*)$ (especially near $g=1/2$, where an unusual first-order transition occurs at $T=0$) are
so weakly first-order (unless they are of some more exotic continuous kind, which cannot be completely
ruled out).
 
The CMF calculations indicate that there may be a narrow region
of first-order transitions for $g<1/2$ (for the ferromagnetic--paramagnetic transition). It is not
clear whether this is an artifact of the small cluster size (with the first-order behavior
obtaining for a $4\times 4$ cluster but not for $2\times 2$). Further large-scale MC simulations and TM
calculations in this region are called for.

\begin{acknowledgments}

We would like to thank Andreas Honecker and Ansgar Kaltz for several stimulating discussions. WG also thanks C.-X. Ding for valuable 
discussions. This research was supported by the NSF under Grants No.~DMR-1104708 and PHY-1211284 (AWS) and by the NSFC under 
Grant 11175018 (WG). WG also gratefully acknowledges the hospitality and financial support from the Condensed Matter 
Theory Visitors Program at Boston University.

\end{acknowledgments}

\end{document}